\documentclass[twocolumn,showpacs]{revtex4}
\usepackage{graphicx}
\usepackage{dcolumn}

\def\be{\begin{equation}}
\def\ee{\end{equation}}
\def\bea{\begin{eqnarray}}
\def\eea{\end{eqnarray}}

\begin{document}
\input epsf
\draft
\renewcommand{\topfraction}{0.8}
\preprint{astro-ph/0301087, \today}
\title {\Large\bf Dark Energy and the Fate of the Universe  }
 \author{\bf Renata Kallosh and Andrei Linde}
\affiliation{ { Department
  of Physics, Stanford University, Stanford, CA 94305-4060,
USA}    }

{\begin{abstract}
It is often assumed that in the course of the evolution of the universe, the dark energy  either vanishes or becomes a positive constant. However, recently it was shown that in many models based on supergravity, the dark energy eventually becomes negative and the universe collapses within the time comparable to the present age of the universe. We will show that this conclusion is not limited to the models based on supergravity: In many models describing the present stage of acceleration of the universe, the dark energy eventually becomes negative, which triggers the collapse of the universe  within the time  $t = 10^{10}-10^{11}$ years. The theories of this type have certain distinguishing features that can be tested by cosmological observations. 
\end{abstract}}
\maketitle 

\section{Introduction}

Recent observations indicate that the Universe is accelerating, and it is spatially flat ($\Omega_{\rm tot}= \Omega_{M}+\Omega_D = 1$). Approximately $0.3$ of the total energy density of the universe $\rho_0 \sim 10^{-120} M_p^4 \sim 10^{-29}$ g/cm$^{3}$ consists of ordinary  matter ($\Omega_{M} \approx 0.3$), and  $0.7$ of the energy density corresponds to dark energy ($\Omega_D \approx 0.7$) \cite{supernova,Bond}. There are three basic scenarios describing the evolution of the universe filled by dark energy.

i) From a purely phenomenological point of view, the simplest possibility is that the dark energy $\rho_D$ is represented by a positive  vacuum energy (cosmological constant) $\Lambda \sim 0.7\rho_0$. If this is the case, the universe will reach de Sitter (dS) regime and expand exponentially for an indefinitely long time,  $a\sim e^{Ht}$. This possibility seems much more natural than the other two possibility to be discussed below. 

ii) It may also happen that the dark energy is the energy of a slowly changing scalar field $\phi$ with equation of state $p_D = w\, \rho_D$,\, $w  \approx -1$.  In most of the models of dark energy it is assumed that the cosmological constant is equal to zero, and the potential energy $V(\phi)$ of the scalar field  driving the present stage of acceleration, slowly decreases and eventually vanishes as the field rolls to $\phi =\infty$, see e.g.  \cite{Dolgov:gh}. In this case, after a transient dS-like stage, the speed of expansion of the universe decreases, and the universe reaches  Minkowski regime.

iii) It is also possible that $V(\phi)$ has a minimum at $V(\phi) <0$, or that it does not have any minimum at all and the field $\phi$ is free to fall to $V(\phi) = -\infty$. In this case {\it the universe eventually collapses,  even if it is flat} \cite{Linde1986}.
 The simplest way to understand this unusual effect is to analyse the Friedmann equation $\left({\dot a\over a}\right)^2  =\rho/3$ (in units $M_p =1$). The positive energy density of a normal matter, as well as the positive kinetic energy density of the scalar field, tend to decrease in an expanding universe. At some moment, the total energy density $\rho$, including the negative contribution $V(\phi) <0$,  vanishes. Once it happens,  the universe, in accordance with the equation $\left({\dot a\over a}\right)^2  =\rho/3$, stops expanding and enters the stage of irreversible collapse.

The last possibility for a while did not attract much attention.  There was no specific reason to expect that the present regime of acceleration is going to end, and there was even less reason to believe that the universe is going to collapse any time soon.

Unfortunately, despite many attempts, we were unable to obtain a good theoretical description of the models i) and ii) in the context of M-theory. We will discuss this issue in Sect. 2. Meanwhile, in \cite{Kallosh:2002gf,Kallosh:2002gg} it was found that one can describe the present state of acceleration of the universe in a broad class of models based on  N=8 extended supergravity (the theory closely related to M-theory). However, the universe described by these models typically collapses within the time $t_{\rm collapse}$ comparable to its present age $t_0 \sim 14$ billion years.

In the beginning, this seemed to be a model-specific result that should be taken seriously only if one can construct fully realistic models of elementary particles based on extended supergravity. However,  this result is valid not only in for the models based on extended supergravity but for many phenomenological models of dark energy based on the minimal N=1 supergravity \cite{Kallosh:2002gf}. 

In this paper we will argue that this result is even more general. It can be obtained in almost every model of dark energy, either based on supergravity or not, if one takes into account the possibility that the effective potential of the field $\phi$ may have a minimum at $V(\phi) <0$ or may be unbounded from below.

We will first review the possibilities to describe the accelerating universe starting with M/string theory and supergravity. Then we will describe our general argument.

\section{Dark energy in M-theory and extended supergravity}

\subsection{M-theory with compactification}

The standard approach to the description of our world in M/string-theory is based on the assumption that our space-time is 11 or 10 dimensional, but 7 or 6 of these dimensions are compactified. The finitness of the volume of the compactified space is required so that   the original D=11 or D=10 theory can be related to the resulting D=4 theory. One can describe the scale of compactification for example, in string theory by introducing scalar moduli field $\phi$, which appears as a coefficient $e^{-\sqrt 6 \phi}$ in front of the potential energy $V$ in 4D. (We use the units $M_p = 1$, where $M_p^2 = {1\over 8\pi G}$.) Here  $V$ is the value of the potential at which other scalar fields of the higher-dimensional theory are stabilized. 
However, the mechanism of stabilization of the compactified space is still unknown. In the absence of such mechanism, the term $e^{-\sqrt 6 \phi}\ V$ leads to the runaway behaviour $\phi\to \infty$, which implies {\it decompactification}. During this process, the energy density $e^{-\sqrt 6 \phi}\ V$ falls down very quickly. A similar result is valid if we consider D=10 string theory as a result of compactification of D=11 supergravity. 
The bottom line is: {\it during the cosmological evolution in D=4 the runaway   moduli represent decompactification of all internal dimensions of M-theory}.

The scale factor of the universe $a(t)$ in the theories with exponential potentials of the type of  $e^{-\lambda \phi}\ V$ grows as $t^{2/\lambda^2}$. In order to describe acceleration of the universe in such theories one would need to have  $\lambda < \sqrt 2$. In the theory with the potential $e^{-\sqrt 6 \phi} V$ the universe can only decelerate. 

In application to string cosmology these observations imply that until we learn how to stabilize the compactified space, we cannot describe the accelerating universe approaching dS regime, as well as the accelerating universe with the energy density slowly approaching zero.   

\subsection{M-theory with non-compactification} 

One way to avoid the problems discussed above is to consider the models with a non-compact internal space  \cite{Kallosh:2001gr,Kallosh:2002wj}.
One may start with D=11 or D=10 supergravity with internal space with an infinite volume and relate it to N=8 supergravity in D=4   \cite{Hull:1988jw}. This is called `non-compactification.' In this approach the connection between the original D=10 or D=11 theory and our D=4 world is more complicated than in the usual case of dimensional reduction. However, one can study these theories directly in D=4. These theories are interesting because they have  maximal amount of supersymmetry,  related to D=11 and/or D=10 supergravities with non-compact internal spaces. 

The number of such models successfully describing an accelerated universe is very limited, due to the maximal amount of supersymmetries. Some of these theories have  dS solutions and can describe dark energy \cite{Kallosh:2002gf,Kallosh:2002wj}.  
These dS solutions correspond to the extrema of the effective potentials $V(\phi)$ for some scalar fields $\phi$. An interesting and very unusual feature of  these scalars in all known theories with $N\geq 2$ is that their mass squared is quantized in units of the Hubble constant $H_0$ corresponding to dS solutions:
${m^2\over H_0^2}=n $,
where $n$ are some integers of the order 1. 
This property was first observed in \cite{Kallosh:2001gr} for a large class of extended supergravities with unstable dS vacua, and confirmed and discussed in detail more recently  in \cite{Kallosh:2002wj} with respect to a new class of $N=2$ gauged supergravities with stable dS vacua \cite{Fre:2002pd}. 

The meaning of this result can be explained in the following way. The simplest potential for a scalar field $\phi$ in this theory has the form
\be \label{simplepot0}
V(\phi) = \Lambda\left(2-\cosh\sqrt 2\phi\right) \ .
\ee
 Usually  the potential near its extremum can be represented as $V(\phi) = \Lambda + m^2\phi^2/2$, where $\Lambda$ and $m^2=V''(0)$ are two free independent parameters. However, in extended supergravities with $\Lambda > 0$ one always has $m^2= V''(0)= n V(0)/3=n \Lambda/3$, where $n$ are integers (we are using units $M_p = 1$) \cite{Kallosh:2001gr,Kallosh:2002wj}. Taking into account that in dS space the Hubble constant is given by $H_0^2 = \Lambda/3$, one has, for $|\phi| \ll 1$, $ 
V(\phi) = \Lambda(1 + n \phi^2/6) = 3H_0^2(1 + n \phi^2/6)$.
In particular, in all known versions of $N=8$ supergravity dS vacuum corresponds to an unstable maximum, $m^2 = -6H^2$ \cite{Kallosh:2001gr,Kallosh:2002wj}, {\it i.e.} at $|\phi| \ll 1$ one has
\be \label{simplepot}
V(\phi) = \Lambda(1 -\phi^2) = 3H^2(1 -\phi^2) \ .
\ee
One can easily verify that the simplest potential (\ref{simplepot0}) satisfies this rule. The main property of this potential is that $m^2 = V''(0) = -2V(0)=-2\Lambda= -6H_0^2$. One can show that a homogeneous field $\phi \ll 1$ with $m^2 = -6H_0^2$ in the universe with the Hubble constant $H_0$ grows as follows: $\phi(t) = \phi_0 \exp{c H_0 t}$, where $c = (\sqrt{33}-3)/2\approx 1.4$. Consequently, in the universe with the energy density dominated by $V(\phi)$ it takes time $t\sim   0.7 H_0^{-1} \ln\phi_0^{-1}$ until the scalar field  rolls down from $\phi_0$ to the region $\phi \gg 1$,  where $V(\phi)$ becomes negative. Once it happens, the universe rapidly collapses.

 \begin{figure}[h!]
\centering\leavevmode\epsfysize=5.3cm \epsfbox{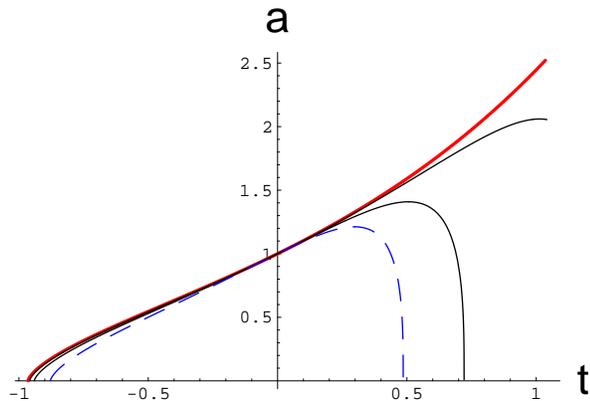}

\

\caption[fig1]
{Scale factor $a(t)$ in the model based on $N=8$ supergravity. The upper (red) curve corresponds to the model with $\phi_0 = 0$. In this case the universe can stay at the top of the effective potential for an extremely long time, until it becomes destabilized by quantum effects \cite{Kallosh:2001gr}. The curves below it correspond to $\phi_0 = 0.2$ and $\phi_0 = 0.3$. The blue dashed curve corresponds to $\phi_{0} = 0.35$.  The present moment is $t=0$. Time is given in units of $H^{-1}(t=0) \approx 14$ billion  years.}
\label{ScalefactorColl}
\end{figure}

Note, that the present age of the universe $t_0$ is approximately equal to $H_0^{-1}$, and the total time of the development of the instability leading to the global collapse of the universe is given by $t\sim   0.7 H_0^{-1} \ln\phi_0^{-1}$, which is also of the same order as $H_0$, unless $\phi_0$ is exponentially small. This explains the main result of Ref. \cite{Kallosh:2002gf,Kallosh:2002gg}:  the universe described by this class of theories is going to collapse within  the time $t_{\rm collapse}$ comparable to its present age $t_0 \sim 14$ billion years, see Fig. \ref{ScalefactorColl}.

\subsection{Two-field models of dark energy in N=8 supergravity}

There are also 3 different cosmological models of dark energy based on N=8 supergravity, with the following potentials of two fields, $\phi$ and $\sigma$. We will list them here, leaving a detailed description of their cosmological implications  for the future publication \cite{Shmakova}
\be \label{332}
V(\phi,\sigma) = \Lambda e^{-\sqrt {2\over 3}\sigma}  \left(3- \cosh (2\sqrt {2}\phi)\right)  \ .
\ee
This is an improved version of Townsend's model of M-theory quintessence \cite{Townsend:2001ea}. Townsend used this potential at $\phi=0$ and found the exponential potential  $e^{-\sqrt {2\over 3}\sigma}$
which may describe the current acceleration with $a(t) \sim t^{3}$. The complete form of the potential shows that the full potential of this theory is unbounded from below and unstable with respect to the generation of the field $s$. The speed of the development of the instability is determined by the curvature of the potential in the $\phi$-direction, $V'' =-4V(0)$, i.e. $m^2_\phi \approx -12 H_0^2$. This allows for the existence of a stage of accelerated expansion, but eventually the universe collapses, just like in the theory (\ref{simplepot0}) discussed above. 

The second model is
\be \label{332a}
V(\phi) = \Lambda e^{-\sqrt {2\over 7}\sigma}  \left(24 e^{-\sqrt {1\over 21}\phi} -8 e^{2\sqrt {3\over 7}\phi}- 3 e^{-8\sqrt {1\over 21}\phi}\right)  \ .
\ee
The $\phi$ part of the potential has a dS maximum at  $\phi=- \sqrt {3\over 7} \log 2$. It may describe the current acceleration during the slow growth of the field $t$: \ $a(t) \sim t^{7}$.  The instability with respect to the field $\phi$ eventually develops and the universe collapses.

The last model of this type has a saddle point dS solution at $\phi=0,\ \sigma=0$
\be \label{332b}
V(\phi) = \Lambda \left(2\cosh \sigma- \cosh \sqrt 2 \phi\right)   \ .
\ee
At the point $\sigma=0$ this model reduces to the model (\ref{simplepot0}) which describes dark energy and eventual collapse of the universe.

A lot of work should be done to incorporate usual matter fields and construct realistic cosmology in N=8 supergravity. However, it is quite encouraging  that there is a class of models based on N=8 supergravity which can describe the present stage of acceleration of the universe. All of these models share the same property: at some moment expansion of the universe stops and the universe collapses within time $t_{\rm collapse}$ comparable to its present age $t_0 \sim 14$ billion years.

\section{A general class of dark energy models}

The main reason for the coincidence of the two different time scales, the present age of the universe and the time until the Big Crunch,  is the relation $|m^2| \sim H^2$. But this
relation   appears not only in the extended
supergravity. It is often valid for the moduli fields  in $N=1$ supergravity 
\cite{Dine:1983ys}. In particular, it is valid in
the simplest Pol\'{o}nyi-type toy model for dark energy in $N=1$
supergravity \cite{Kallosh:2002gf}.  If one does not fine-tune the value of the cosmological constant in this model to be equal to zero, one has two equally compelling options: $V(\phi) >0$ and $V(\phi) <0$ in the minimum of the effective potential. In the first case, the universe enters dS regime of eternal expansion. In the second case, the universe collapses. The time until the global collapse depends on the parameters of the model and the initial conditions, but typically it is of the same order as $t_0 \sim 14$ billion years, just as in the extended supergravity \cite{Kallosh:2002gf}.

Another interesting model  is the axion quintessence  \cite{Frieman:1995pm,Choi:1999xn}. In the M-theory motivated version of this model proposed in \cite{Choi:1999xn} one has $V(\phi) \sim \Lambda (\cos(\phi/f)+C)$, where the value of constant $C$   depends on the details of the model. For $C=0$, $f = O(M_p)$ one finds $m^2 = V''(0) = -O(H^2_0)$. According to \cite{Kallosh:2002gf,Jan}, this version of the axion quintessence model can successfully describe the present stage of acceleration of the universe, but, just like the $N=8$ models, it leads to a global collapse of the universe in the future within the typical time $t_{\rm collapse}\sim t_0 \sim 14$ billion years.

In fact, the crucial relation $|m^2| \lesssim H^2$ is valid in most of the dark energy models. It is the standard inflationary slow-roll condition, which should be valid at the present stage of the late-time inflation/acceleration of the universe. Sometimes this slow-roll condition can be violated \cite{Kallosh:2001gr,Linde:2001ae}, but typically one cannot have a prolonged stage of acceleration of the universe for $|m^2| \gg H^2$. As a result, one can give a simple  argument suggesting  that the coincidence of the two different time scales, the present age of the universe and the time remaining until the Big Crunch, is a generic property of many models of dark energy.

Let us make the simplest assumption that the expansion of the universe at $t > t_0$ can be approximately described by the simple power-law equation 
$  
a(t) \sim a(t_0)\left({t+c\over t_0+c}\right)^r,
$
where $c$ is some constant.
The Hubble constant in the universe with $a(t) \sim (t+c)^r$ is given by ${r\over t+c}$, which means that the total energy density is equal to $3H^2= {3r^2 \over (t+c)^2}$. The parameter $c$ is determined by the continuity condition for $H$:
\ $ 
{r\over t_0+c}=H_0 \approx t_0^{-1}.$ 
The last part of this equation is the observed relation between the present value of $H$ and the age of the universe in the simplest $\Lambda$CDM model. This gives $c\approx (r-1)t_0$.
 
The energy density will become $N^2$ times smaller than now at the time $t-t_0 \approx (N-1) \, r\, t_0$. In particular, the energy density will become 2 times smaller at the time $t-t_0\approx 0.4\, r\, t_0$ from now, and it will become 9 times smaller at $t-t_0=  2r\, t_0$. How large is this time interval?

 Acceleration of the universe implies that $r> 1$. 
Taking for definiteness $r=2$, one finds that the density of the universe will decrease 2 times at the moment  $t-t_0\approx 0.8\, t_0\sim 10$  billion years from now, and it will drop 9  times $50$ billion years from now.

Now let us take any model of dark energy with $r =2$ and add to the scalar potential $V(\phi)$ a tiny negative cosmological constant $\Lambda =-\rho_0/N^2$ with $N^2\gg 1$. The evolution of the universe up to the present moment will not change significantly.  One can make this model even better by multiplying $V(\phi)$ by some factor $B> 1$ to ensure that the present value of $\Omega_D$ remains equal $0.7$ even after we add a negative constant $\Lambda =-\rho_0/N^2$ to $V(\phi)$. This model will remain a viable model of dark energy. However, after some time $t-t_0\approx 2(N-1) \, t_0$ the value of $V(\phi)$ will drop down more than N times, the total energy density of the universe will vanish, the universe will stop expanding, and soon after that it will collapse.  If $N$ is not too large,  the universe collapses at the time $t_{\rm collapse} = O(t_0)$.

One could think that the argument given above works only for the marginal situation, when $a(t) \sim t^r$ with $r = O(1)$. However, the final result is very general. Our procedure of modification of the potential (subtraction of a  constant and a subsequent compensation of the decrease of $\Omega_D$ via the multiplication of $V(\phi)$ by a constant $B> 1$) works until the moment when the theory no longer represents dark energy because the potential becomes too steep. This happens for any theory of dark energy, even if the original potential was extremely flat (which corresponds to  $r\gg 1$). Once it happens, the field rolls down within the time  $O(H^{-1})$. The rolling field rapidly approaches the region of negative $V(\phi)$, which typically leads to the collapse of the universe within the time  $O(H^{-1})$.

This suggests that many  models of dark energy considered in the literature have viable counterparts that can be obtained from the original models by adding a  negative cosmological constant $\Lambda \lesssim -\rho_0$. Unless the absolute value of this extra term is many orders of magnitude smaller than $\rho_0$, these models will describe the universe collapsing within the time comparable to the present age of the universe.

\section{Example: Dark energy with an exponential potential}

As an example illustrating the general argument given in the previous section, let us consider dark energy described by the scalar field $\phi$ with an exponential potential
\begin{equation}  
V(\phi)= \Lambda\ e^{-\lambda \phi}.
\end{equation}
We already mentioned that this theory describes an accelerating universe for $\lambda < \sqrt 2$. However, in the realistic situation this condition can be slightly relaxed. One should take into account that the post-inflationary universe rapidly expanded, being dominated by hot matter and  then by cold dark matter. This expansion does not allow  the field $\phi$ to move until the energy density of matter becomes sufficiently small. Therefore in the beginning the kinetic energy of the scalar field is very small, so it has the same equation of state as the cosmological constant. 
Then the field starts moving slowly, loosing its energy at a much slower paste than CDM. Eventually, the universe enters the stage with $\Omega_D=0.7$ (the present time), and then $\Omega_D$ continues to grow. As a result,  one can have  $\Omega_D =0.7$ and  $w <-0.6$ at present for  $\lambda \lesssim 1.7$ \cite{LopesFranca:2002ek,Kallosh:2002gf}.

In our investigation we will assume, without loosing the generality, that the initial value of the field was $\phi =0$ (one can  always rescale the field and the potential). Then one should find such value of the parameter $\Lambda$ that the universe enters the stage with $\Omega_D=0.7$ at the same time when its Hubble constant acquires its present value $H_0 \sim 10^{-60} M_p$. This requires fine-tuning, but this is the same fine-tuning that hampers all models of dark energy \cite{LopesFranca:2002ek,Kallosh:2002gf}.

  \begin{figure}[h!]
\centering\leavevmode\epsfysize=5cm \epsfbox{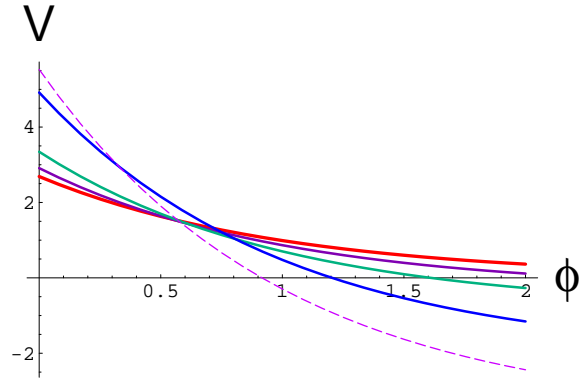}

\

\caption[fig1]
{Effective potential $V = \Lambda_C\left(e^{\phi/2}-C\right)$ with $C=   0,\,  0.1\ ,0.2,\, 0.3 $ and $0.4$. The coefficients $\Lambda_C$ are fixed by the condition that for each value  of $C$ one should have the same value of the Hubble constant and 
$\Omega_D=0.7$ at the present moment $t=t_0$. }
\label{pot}
\end{figure}

Now we will consider a class of potentials of a more general  type, containing a constant negative contribution, as suggested in the previous section:  
\begin{equation} \label{expconst} 
V(\phi)= \Lambda_C\ \left(e^{-\lambda \phi}-C\right).
\end{equation}
Here $C$ is some positive constant. The constant $\Lambda_c$ should be found anew for each new value of $C$, just as we did for the case $C=0$ described above. This can be done, and the cosmological evolution of this model can be easily studied using the methods described in \cite{Kallosh:2002gf}. Here we will only present the results of our investigation of a model with $\lambda = 1$.

 \begin{figure}[h!]
\centering\leavevmode\epsfysize=5.3cm \epsfbox{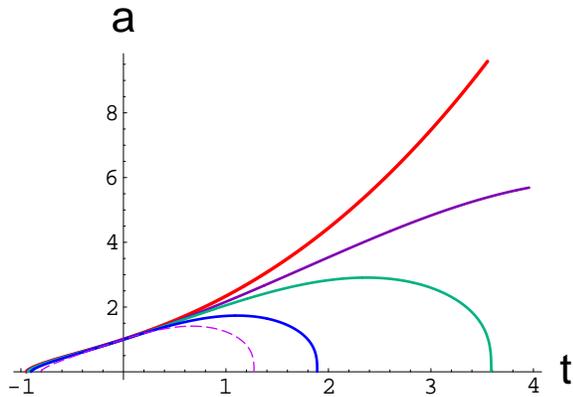}

\

\caption[fig1]
{Scale factor $a(t)$ in the model with the potential $V = \Lambda_C\left(e^{\phi/2}-C\right)$. The upper (red) curve corresponds to the model with $C = 0$. The curves below it correspond to $C=  0.1,\, 0.2,\, 0.3$ and $0.4$. The present moment is $t=0$. Time is given in units of $H^{-1}(t=0) \approx 14$ billion  years.}
\label{ScalefactorColl2}
\end{figure}

Figure \ref{ScalefactorColl2} shows the evolution of the scale factor of the universe for 4 different values of parameter $C$: ~$C=   0,\,  0.1,\, 0.2,\, 0.3$ and $0.4$. For all of these cases one can find such parameters $\Lambda_C$ that the value of the Hubble constant at $t =t_0$ coincides with its present value (the curves $a(t)$ have the same derivative at $t=0$ in Figure \ref{ScalefactorColl2}).

  \begin{figure}[h!]
\centering\leavevmode\epsfysize=5.3cm \epsfbox{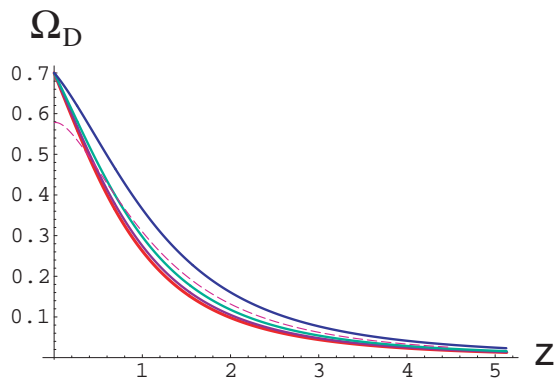}

\

\caption[fig1]
{Dark energy $ \Omega_D$ as a function of redshift $z$ for $V = \Lambda_C\left(e^{\phi/2}-C\right)$ with $C=   0,\,  0.1,\, 0.2,\, 0.3$ and $0.4$. The present time corresponds to $z=0$. As we see, all curves are practically indistinguishable, except for the dashed curve corresponding to $C= 0.3$.}
\label{OmegaColl}
\end{figure}

 \begin{figure}[h!]
\centering\leavevmode\epsfysize=5cm \epsfbox{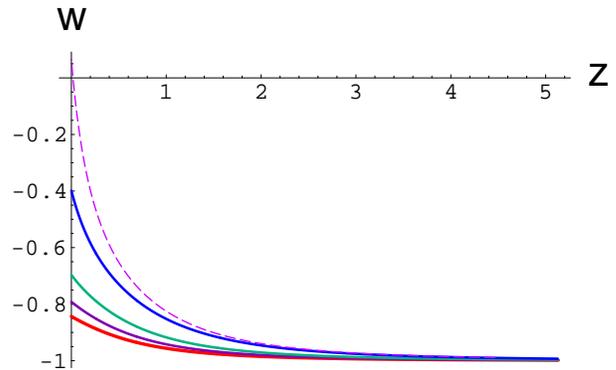}

\

\caption[fig1]
{Equation of state $w$ as a function of redshift $z$ for $C=   0,\,  0.1,\,0.2,\, 0.3 $ and $0.4$.  For  $C=    0.4$ this function sharply rises to $w >0$ near $z=0$. The red (thick) line $w =-1$ corresponds to the model with $C = 0$.}
\label{wColl}
\end{figure}

All of the models with $C=   0,\,  0.1,\, 0.2,\, 0.3$  with the potentials shown in Figure \ref{pot} can represent dark energy with $\Omega_D =0.7$ at the present moment (the point $z=0$ in Figure \ref{OmegaColl}). However, the largest value of $\Omega_D$ in the case  $C=0.4$ is $0.58$, and the equation of state $w$ in this case blows up near $z=0$, see Figure \ref{wColl}.

 \begin{figure}[h!]
\centering\leavevmode\epsfysize=5.7cm \epsfbox{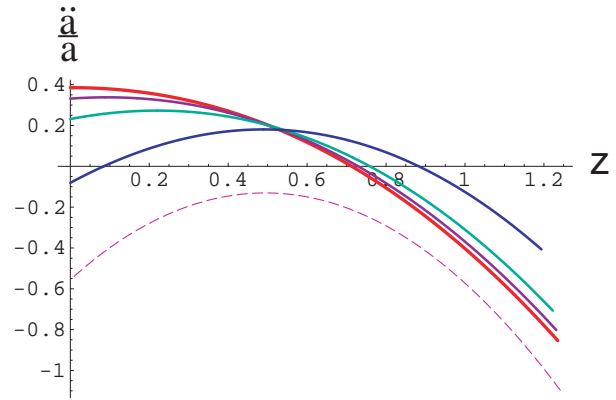}

\

\caption[fig1]
{Relative acceleration of the universe ${\ddot a/a}$ as a function of redshift $z$ for $C=   0,\,  0.1\ ,0.2,\, 0.3 ,$ and $0.4$.  For  $C=    0.4$ the universe never accelerates, which is ruled out by observational data. The red (thick) line $w =-1$ corresponds to the model with $C = 0$.}
\label{acc}
\end{figure}

As we see, the universe with $C=0$ (thick red line in Figures 
\ref{pot}-\ref{acc}) always continues its accelerated expansion. However, in all other cases the universe collapses within the time ranging from $1.3 t_0 \sim 18$ billion years (for $C=0.4$) to $7.5 t_0 \sim 105$ billion years (for $C=0.1$), in agreement with the argument given in the previous section. The universe with $C=0.4$ never accelerates, so this model is ruled out by the existing observations.

Thus, for every model  $V(\phi)= \Lambda\ e^{-\phi}$ successfully describing dark energy there exist many other models $V(\phi)= \Lambda_C\ \left(e^{-\phi}-C\right)$ with $0<C\lesssim 0.3$ which provide an equally good description of the present stage of acceleration, but lead to the global collapse of the universe within the next  $10^{10}-10^{11}$ years.

Similar results can be obtained for many other models of dark energy, e.g. for the theories $V(\phi)= \Lambda_C\ \left(e^{-\lambda \phi}-C\right)$ with all possible values of $\lambda$, or for the models with the inverse power law potentials.

\section{Discussion}

Recent discovery of acceleration of the universe is one of the major challenges for the modern theory of fundamental interactions. After many attempts to explain why the cosmological constant must be zero, the theorists switched to the new paradigm and started trying to explain why it should be positive and why, consequently, the universe should expand forever. The first attempts to do so in the context of M/string theory and extended supergravity revealed many problems described, e.g., in \cite{Gibbons:85,Hellerman:2001yi}. Then we learned that one can describe acceleration of the universe in extended supergravity, but in all models based on extended supergravity, with the exception of the N=2 model of \cite{Fre:2002pd}, the regime of acceleration is unstable \cite{Kallosh:2001gr}. Typically it ends by a global collapse of the universe within the time comparable with the present age of the universe, $t_{\rm collapse} \sim 10-30$ billion years \cite{Kallosh:2002gf,Kallosh:2002gg}. 

In this paper we have shown that the possibility of a global collapse is not specific to supergravity but is, in fact, quite generic. For every model of dark energy describing eternally expanding universe one can construct many closely related models which describe the present stage of acceleration of the universe followed by its global collapse.

This does not mean that we are making a doomsday prediction. None of the existing theoretical models of dark energy look particularly natural and attractive. It may happen that eventually we will find good theoretical models describing an eternally accelerating universe, or conclude, on the basis of anthropic considerations, that dark energy should change in time extremely slowly, so that the collapse will occur exponentially far away in the future \cite{Vil2002}. We hope to return to this question in the future publications.  However, in the absence of a compelling theory of dark energy  one may also consider a more humble approach and try  to compare predictions of various models of dark energy with observations, see e.g. \cite{Copeland}.

If the universe is going to collapse, then in the beginning of this process the speed of expansion of the universe should gradually slow down, see Fig. \ref{acc}. This is accompanied by the rapid growth of the parameter $w$ at small $z$, as shown in Fig. \ref{wColl}. We find it quite significant that some of the models predicting global collapse  can be already ruled out by the existing observational data. For example, all models $\Lambda_C  \left(e^{-\phi}-C\right)$ predicting the global collapse within the next 18 billion years, do not describe the present stage of acceleration, and therefore contradict the recent cosmological observations. 
Thus, even though the observational data cannot rule out the general possibility of the global collapse in the distant future, they can help us to put strong constraints on the time of the possible Big Crunch.

It is a pleasure to thank R. Bond, L. Kofman, J. Kratochvil,  E. Linder, S.~Prokushkin and M.~Shmakova,  for useful discussions. This work was supported by NSF grant PHY-9870115. The work by A.L. was also supported
by the Templeton Foundation grant No. 938-COS273.

\end{document}